\documentclass[prb,twocolumn,showpacs,superscriptaddress]{revtex4}

\usepackage{amsmath}
\usepackage{amssymb}
\usepackage{graphicx}
\usepackage[hypertex]{hyperref}
\usepackage{srctex}
\usepackage[usenames]{color}

\newcommand{\GL}{\mathrm{\scriptscriptstyle GL}}
\newcommand{\Gs}{{\scriptscriptstyle G}}
\newcommand{\sigmaD}{\sigma_{\scriptscriptstyle D}}
\newcommand{\DOS}{\mathrm{\scriptscriptstyle DOS}}
\newcommand{\MT}{\mathrm{\scriptscriptstyle MT}}
\newcommand{\AL}{\mathrm{\scriptscriptstyle AL}}
\newcommand{\cp}{\mathrm{ cp}}

\definecolor{MidnightBlue}{cmyk}{0.98,0.13,0,0.43}
\definecolor{DarkGreen}{rgb}{0,0.7,0.1}

   \DeclareMathOperator{\ch}{ch} 

\begin{document}
\newcommand{\remark}[1] {\noindent\framebox{
\begin{minipage}{0.96\columnwidth}\textbf{\textit{ #1}}
\end{minipage}}
}

\newcommand{\bnabla}{{\boldsymbol{\nabla}}} \newcommand{\Tr}{\mathrm{Tr}} \newcommand{\Dk}{\check{\Delta}_{\cal K}} \newcommand{\Qk}{\check{Q}_{\cal K}} \newcommand{\Fk}{\check{\Phi}_{\cal K}}
\newcommand{\fk}{\check{\phi}_{\cal K}} \newcommand{\Ak}{\check{\mathbf{A}}_{\cal K}} \newcommand{\Si}{\check{\Xi}}
\newcommand{\cK}{\cal K}
\newcommand{\Lk}{\check{\Lambda}}
\newcommand{\bz}{{\mathbf z}}
\newcommand{\bx}{{\mathbf x}}
\newcommand{\br}{{\mathbf r}}
\newcommand{\bG}{{\mathbf G}}
\newcommand{\bu}{{\mathbf u}}
\newcommand{\bq}{{\mathbf q}}
\newcommand{\cH}{{\cal H}}
\newcommand{\dif}{{\mathrm d}}

\def\Xint#1{\mathchoice
{\XXint\displaystyle\textstyle{#1}}%
{\XXint\textstyle\scriptstyle{#1}}%
{\XXint\scriptstyle\scriptscriptstyle{#1}}%
{\XXint\scriptscriptstyle\scriptscriptstyle{#1}}%
\!\int}
\def\XXint#1#2#3{{\setbox0=\hbox{$#1{#2#3}{\int}$}
\vcenter{\hbox{$#2#3$}}\kern-.5\wd0}}
\def\ddashint{\Xint=}
\def\dashint{\Xint-}

\title{Far-from-equilibrium superconductor in fluctuational regime}
\author{A.~Petkovi\'{c}}
\affiliation{Laboratoire de Physique Th\'{e}orique-CNRS, Ecole Normale Sup\'{e}rieure, 24 rue Lhomond, 75005 Paris, France}
\author{N.\,M.~Chtchelkatchev}
\affiliation{Materials Science Division, Argonne National Laboratory, Argonne, Illinois 60439, USA}
\affiliation{Institute for High Pressure Physics, Russian Academy of Sciences, Troitsk 142190, Moscow region, Russia}
\affiliation{L.D. Landau Institute for Theoretical Physics, Russian Academy of Sciences,
Moscow 117940, Russia}
\author{V.\,M.~Vinokur}
\affiliation{Materials Science Division, Argonne National Laboratory, Argonne, Illinois 60439, USA}

\date{\today}

\begin{abstract}
We derive Ginzburg-Landau-like action for two-dimensional disordered superconductor under far-from-equilibrium conditions in a fluctuational regime. Then, utilizing it, we calculate fluctuation induced density of states, Maki-Thomson and Aslamazov-Larkin type contributions to the in-plane electrical conductivity. We apply our approach to thin superconducting film sandwiched between a gate and a substrate that have different temperatures and different electrochemical potentials.
\end{abstract}



\maketitle

\section{Introduction}

Most of processes in physics and in technological realm occur under far from the equilibrium (FFE) conditions.
At the same time, the theories of nonequilibrium behavior were mostly restricted to small deviation from the equilibrium.  A marked progress in approaches to quantitative description of FFE physics is related to Keldysh technique-based methods \cite{Keldysh}.
However, mostly FFE systems have been studied within the nonequilibrium form of the mean-field theory, see, for example,\cite{Kramer,Stoof}.
In the present paper, we address the important question of what happens to the conventional second-order phase
transition if the system under consideration is driven out of equilibrium.
In equilibrium when close enough to the second order phase transition the mean field theory does not hold and the physics starts to be governed by
fluctuations.~\cite{pat,Varlamovbook}
In our work, we construct a theory of fluctuations
under FFE conditions.  As the exemplary system, we consider a superconductor FFE in the fluctuational regime.

Our first notion is that while in clean three dimensional conventional superconductors the fluctuations are important only in a very narrow region
around the superconducting transition line (usually within the $\sim 10^{-12}$K temperature range), in high temperature, low-dimensional
and organic superconductors the fluctuation region is much wider.
In particular, as early as in 1968, Aslamazov and Larkin, and, independently, Maki showed that in disordered thin superconducting films
the width of the fluctuation region, which is determined by the sheet resistance, grows noticeably as compared to that
of bulk superconductors.~\cite{Aslamazov1,Aslamazov2,Maki,Varlamovbook}
Moreover, it was demonstrated that not only thermodynamic but also dynamic characteristics of the low-dimensional systems
are strongly influenced by the fluctuations close to an equilibrium, see e.g. Refs.~\onlinecite{Varlamovbook,L_Kamenev} for a review.

\begin{figure}[hbt]
\includegraphics[width=0.75\columnwidth]{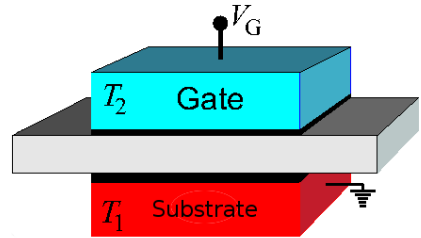}\\
\caption{Thin superconducting film sandwiched between the substrate and the gate. The substrate temperature is  $T_1$, the gate temperature is $T_2$ and the gate voltage is $V_{\Gs}$. Changing $T_{1}$, $T_{2}$ and $V_{\Gs}$ one can tune the nonequilibrium quasiparticle distribution in the film. \label{device}}
\end{figure}

A quantitative approach to nonequilibrium fluctuation superconductivity was recently formulated in Refs.~\onlinecite{NikolayEPL,PRL}.
Building on this approach, we develop further our original Keldysh technique enabling us to find  nonequilibrium fluctuation contributions
to the electrical conductivity of a superconductor above the (nonequilibrium) superconducting transition.
We show that by measuring the fluctuation corrections one can  infer the parameters of the nonequilibrium state of the superconductor from the
experimental data.  Further, while in the equilibrium the lifetime of the fluctuation induced Cooper pairs is determined by the difference $T-T_c$,
where $T_c$ is the critical temperature\cite{Varlamovbook,L_Kamenev}, we find that in FFE conditions it is controlled by the
parameters of the nonequilibrium density matrix of the system.
For example, for a thin superconducting film sandwiched between the gate and the substrate,
see Fig.~\ref{device}, these parameters are the temperatures of the gate and of the substrate, and the gate voltage $V_{\Gs}$.

The paper is organized as follows. In Sec.~\ref{sec1} we introduce the model. In Sec.~\ref{sec:third} we derive the nonequlibrium Ginzburg-Landau (GL) like action, while in Sec.~\ref{sec4} we calculate fluctuation induced corrections to the electrical conductivity of the normal metal. In Sec.~\ref{Sec:example} we focus on the specific experimental realization shown in Fig.~\ref{device} and analyze the previously derived results for this concrete setup. In Sec.~\ref{sec:conclusions}, we summarize our results and discuss their further applications. Technical details are relegated to Appendix A.

\section{Keldysh sigma-model action \label{sec1}}

The Keldysh partition function of a superconductor with the Hamiltonian $H$ in the coherent state basis is defined as:
\begin{align}
\mathcal Z=\mathcal{N}\int D\bar{\psi}D\psi \exp\{iS[\bar{\psi},\psi]\},
\end{align}
where
\begin{align}\label{eq:startS}
S[\bar{\psi},\psi]=\int_{\cal C}dt\left\{\int_{\mathbf{r}}(\bar\psi_\alpha i \partial_t{\psi_\alpha})-H[\bar{\psi},\psi]\right\},
\end{align}
and $\mathcal{N}$ is a normalization constant. Hereafter $\hbar=c=k_B=1$.
Here ${\cal C}$ is the Keldysh contour and $\alpha\equiv\uparrow$,$\downarrow$ is the spin variable.

The Hamiltonian has the form $H=H_0+H_{\rm int}$, where the single-particle Hamiltonian
\begin{align}\label{eq:singleH}
H_0=\int_{\mathbf r}\bar\psi_\alpha\left[-\frac{(\bnabla - ie\mathbf{A})^2}{2m} +U_{\rm dis} +e \phi \right]\psi_\alpha.
\end{align}
Here $\mathbf A$, $\phi$ and $U_{\rm dis}$ are vector, scalar and disorder potentials and $e$ is the electron charge;
the tensor summation over the spin indices is implied.
The interaction Hamiltonian describes the electron-electron interaction in the Cooper channel,
\begin{align}
H_{\rm int}=
-\frac{\lambda}{\nu} \int_{\mathbf{r}}  \bar\psi_{\uparrow}\bar\psi_\downarrow\psi_\downarrow \psi_\uparrow\, ,
\end{align}
where $\rho(\mathbf r)=\sum_{\alpha}\bar\psi_\alpha(\mathbf r)\psi_\alpha(\mathbf r)$ is the local electron density and the superconductive coupling constant $\lambda>0$. The disorder potential is assumed to be Gaussian distributed with the correlator
\begin{align}
\langle U_{\mathrm{dis}}(\br)U_{\mathrm{dis}}(\br')\rangle =\frac{1}{2\pi\nu\tau}\delta(\br-\br').
\end{align}
Averaging over disorder and carrying out the standard decoupling in the four-fermion terms in the action via the Stratonovich-Hubbard fields $\{Q,\Delta\}$,~\cite{L_Kamenev}
and integrating out the degrees of freedom with the energies higher than $1/\tau$, where $\tau^{-1}$ is the elastic scattering rate, we arrive at the Keldysh nonlinear $\sigma$-model partition function
\begin{align}
\label{eq:Z}
&Z=\int D[Q,\Delta]\exp\{i S[\Qk,\Dk]\},
\\\label{eq:S}
&S[\Qk,\Dk]=S_{\Delta}+S_{\phi}+S_{Q}.
\end{align}
The nonlinear $\sigma$-model action $S$ consists of three parts:
\begin{align}
S_{\Delta}=&-\frac{\nu}{2\lambda}\Tr[\Dk^{\dagger}\check{Y}\Dk],\\ S_{\phi}=&\frac{e^2\nu}{2}\Tr[\check \phi_{\mathcal{K}}\check{Y}\check \phi_{\mathcal{K}}],\\
\label{q:S_Q}
S_{Q}=&\frac{i \pi \nu}{4}\Tr[D (\partial_{\bf{r}}\Qk)^2-4 \Si\partial_t\Qk\notag\\
&-4ie\check \phi_{\mathcal{K}}\Qk+4i\Dk\Qk ],
\end{align}
where $\check Q^2=1$. Here $D$ is the diffusion coefficient and $\nu$ is the bare single particle density of states at the Fermi level per one spin projection. The action \eqref{eq:S} holds while the effective temperatures (see Sec.~\ref{sec4}) that follow from it are much smaller than $1/\tau$. The check mark above the field variables indicates that they are defined in the space that is the tensor product of the Keldysh and Nambu spaces. The former and later are spanned by the Pauli matrices $\hat{\sigma}_i$ and $\hat{\tau}_i$, $i\in\{0,x,y,z\}$, respectively. So, $\check{Y}=\hat{\sigma}_x\otimes \hat{\tau}_0$, $\Si=\hat{\sigma}_0\otimes\hat{\tau}_z$. Multiplication in time space is implicitly assumed, and ``Tr'' includes an integration over real space. The subscript $\cK$ denotes the gauge transformed fields:
\begin{align}
\check\phi_{\cK}&=\check{\phi}-\partial_{t}\check{\cK},\\ {\Ak}&={\check{\mathbf{A}}} +{\bf{\bnabla}}{\check{\cK}},\\ \check{\cK}&=[k^{cl}\hat{\sigma}_{0}+k^{q}\hat{\sigma}_{x}]\otimes\hat{\tau}_{0}.
\end{align}
$\check A$ and  $\check \phi$ are defined in same way as $\check{\cK}$. Also
\begin{align}
\check{\Delta}&= [\Delta^{cl}\hat{\sigma}_{0}+ \Delta^{q}\hat{\sigma}_{x}]\otimes\hat{\tau}_{+}-\mathrm{h.c.},\\
\Dk(\br,t)&=e^{ie \Si{\check{\cK}}(\br,t)}\check{\Delta}e^{-ie\Si{\check{\cK}}(\br,t)}.
\end{align}
$\Qk$ is defined in the same way.
The quantum (q) and classical (cl) components are defined as half-sum and half-difference of the field values at the lower and upper branches of the Keldysh time-contour. The field $\Delta^{cl}$ becomes the superconducting order parameter on the mean-field (saddle-point) level, while the saddle point equation for $\check Q$ produces the Usadel quasiclassical equations, where $\check Q$ plays the role of the quasiclassical Greens function. The covariant spatial derivative is given by
\begin{align}
\partial_{\mathbf{r}}\check{Q}_{\cK}= \bnabla_{\mathbf{r}}\check{Q}_{\cK}- ie[\check{\Xi}\check{\mathbf{A}}_{\cK},\check{Q}_{\cK}].
\end{align}
\section{Ginzburg-Landau action\label{sec:third}}

First, let us consider simplified situation ignoring the interactions. Then, the metallic saddle point of Eq.(\ref{eq:S}), is \cite{first,second,third}
\begin{align}
\check \Lambda&=\check{\mathcal{U}}\check{\Lambda}_0 \check{\mathcal{U}}^{-1},\quad\quad\check{\Lambda}_0=\hat{\sigma}_z\otimes\hat{\tau}_z,\\
\label{lambda}
\check{\mathcal{U}}_{t,t'}(\br)&=\check{\mathcal{U}}_{t,t'}^{-1}(\br)=
\left(\begin{array}{cc}\delta_{t-t'-0}\hat{\tau}_0&
\hat{F}_{t,t'}(\br)\\
0&-\delta_{t-t'+0}\hat{\tau}_0\end{array}\right),\\
\hat{F}_{t,t'}(\br)&=\begin{pmatrix}
F_{t,t'}^e(\br)& 0\\
0& F_{t,t'}^h(\br)
\end{pmatrix}.
\end{align}
By setting quantum components of electromagnetic potentials to zero, the equation for $F^{e/h}_{t,t'}(\br)$ reads:
\begin{align}
&D\Big(\bnabla^2 F^{e/h}\mp 2ie[{\bf{A}}^{cl}_{\cK},{\bnabla} F^{e/h}]\mp i e [{\bnabla}{{\bf{A}}^{cl}_{\cK}}, F^{e/h}]\notag\\&-\frac{e^2}{2}[{{\bf{A}}^{cl}_{\cK}},[{{\bf{A}}^{cl}_{\cK}},F^{e/h}]]\Big) -\overrightarrow{\partial_t } F^{e/h}+F^{e/h}\overleftarrow{\partial_t}\notag\\&\mp i e[\phi^{cl}_{\cK}, F^{e/h}]=0.
\end{align}
After the Wigner transformation, we can
map $F^{e/h}$ to quasiparticle electron (hole) distribution functions:
$F^{e/h}_{\epsilon}(\br,t)\equiv 1-2f^{e/h}_{\epsilon}(\br,t)$. In the leading order with respect to time and energy partial derivatives of external potentials, the kinetic equations  become
\begin{multline}\label{eq:kinetic}
D\big[ {\bnabla}^2F^{e/h}_{\epsilon}\mp 2 e (\partial_{t}{{\bf{A}}^{cl}_{\cK}})(\partial_{\epsilon}\bnabla F^{e/h}_{\epsilon})
\\
+\frac{1}{2} e^2(\partial_{t}{{\bf{A}}^{cl}_{\cK}})^2(\partial^2_{\epsilon} F^{e/h}_{\epsilon})\big]-\partial_tF^{e/h}_{\epsilon}
\\
\mp e\partial_{t}(D \bnabla{{\bf{A}}^{cl}_{\cK}}+\phi^{cl}_{\cK}) \partial_{\epsilon} F^{e/h}_{\epsilon}=0\,.
\end{multline}
Note that one has to take into  account adequate boundary conditions.

Having specified the metallic saddle point solution, we consider the massless fluctuations around it. They can be parameterized as \cite{Feigelman+00}
\begin{align}\label{parametrization}
\check{Q}_{\cK}(\mathbf{r}) &= e^{- \check{W}(\mathbf{r})/2}\, \check{\Lambda}(\br)\,\, e^{\check{W}(\mathbf{r})/2 },\quad \check{W}=\check{\mathcal{U}}\check{\mathcal{W}} \check{\mathcal{U}}^{-1},\\
\label{eq:W}
\check{\mathcal{W}}&=\left(\begin{array}{cc}
w\tau_{+} - w^{*}\tau_{-} & w_{0}\tau_{0}+w_{z}\tau_{z}\\
\bar{w}_{0}\tau_{0}+\bar{w}_{z}\tau_{z} & \bar{w}\tau_{+} -
\bar{w}^{*}\tau_{-}
\end{array}\right),
\end{align}
such that $\check{W}\check{\Lambda}+\check{\Lambda}\check{W}=0$. Here we introduced four real fields $w^{\alpha}_{tt'}(\mathbf{r}),\bar{w}^{\alpha}_{tt'}(\mathbf{r})$
with $\alpha=0,z$ representing diffuson degrees of freedom  and two complex
fields $w_{tt'}(\mathbf{r}),\bar{w}_{tt'}(\mathbf{r})$ for Cooperon degrees freedom.

Since the main goal of this section is the derivation of the nonequilibrium extension of GL-like action in the normal state but very close to the transition,
in what follows we will concentrate on Cooperon degrees of freedom only.
By plugging $\Qk$, given by Eq.~(\ref{parametrization}), in the action Eq.~(\ref{eq:S}) and expanding it up to the second order in Cooperons we get
\begin{align}\label{eq:S_Q}
S_{Q}&=S_{w^2}
+S_{w\Delta},\\
S_{w^2}&=i\frac{\pi\nu}{2}\Tr\left[ \mathbf{w}_{t,t'}^\dag(\br)\mathcal{C}^{-1}_{t,t'}(\br)\mathbf{w}_{t',t}(\br)\right],\\
S_{w\Delta}&=-\pi\nu \Tr\Big[(-\Delta^{cl*}-F_h \Delta^{q*})w+(\Delta^{cl}+F_e \Delta^{q})w^*\notag\\&+(\Delta^{cl*}-F_e \Delta^{q*})\bar{w}+(-\Delta^{cl}+F_h \Delta^{q})\bar{w}^*\Big].
\end{align}
Here $\mathbf{w}_{t,t'}(\br)=(w_{t,t'}(\br),\bar{w}_{t,t'}(\br))^T$ and \cite{L_Kamenev}
\begin{align}
\mathcal{C}^{-1}_{t,t'}(\br)&=\left(\begin{array}{cc}
C_{t,t'}^{-1}(\br) & 0\\
0 & \bar{C}_{t,t'}^{-1}(\br)
\end{array}\right),\\
C_{t,t'}^{-1}(\br)&=-\partial_t+\partial_{t'}-ie\left[\phi^{cl}_{\cK}(\br,t)- \phi^{cl}_{\cK}(\br,t')\right]\notag\\&-D\left[\bnabla-i e \mathbf{A}^{cl}_{\cK}(\br,t)-i e \mathbf{A}^{cl}_{\cK}(\br,t')\right]^2,\\
\bar{C}_{t,t'}^{-1}(\br)&=\partial_t-\partial_{t'}+ie\left[\phi^{cl}_{\cK}(\br,t)- \phi^{cl}_{\cK}(\br,t')\right]\notag\\&-D\left[\bnabla-i e \mathbf{A}^{cl}_{\cK}(\br,t)-i e \mathbf{A}^{cl}_{\cK}(\br,t')\right]^2.
\end{align}
Note that there are additional terms in the action (\ref{eq:S_Q}), that are not presented here for the sake of brevity.
They contain quantum components of electromagnetic potentials.
Since quantum components of fields are the auxiliary source fields usually used to calculate observables by
the appropriate differentiation of the Keldysh action, we omit the mentioned terms here.
They are not important for the derivation of the GL action.
We will discuss these terms in the next section, when calculating different corrections to the Drude conductivity that arise due to superconducting fluctuations.

Now, the Cooperon degrees of freedom can be integrated out from the Keldysh partition function (\ref{eq:Z}). In this way the GL action is generated:
\begin{align}\label{GL}
i S_{\GL}\left[\Delta^{cl},\Delta^{q}\right]=-\frac{1}{2}\langle S_{w\Delta}^2\rangle_{iS_{w^2}}+i S_{\Delta}.
\end{align}
In what follows, we consider stationary distribution functions ( i.e.~$F_{t,t'}=F_{t-t'}$) and stationary electromagnetic fields.
Also, we assume that they slowly vary in space as compared to the fluctuating order parameter $\Delta(\br,t)$. Then, we obtain (see App.~\ref{app:GL})
\begin{align}\label{GLaction}
S_{\GL}\left[\Delta^{cl},\Delta^{q}\right]&=2\nu \Tr\left[\mathbf{\Delta}_{\cK}^{\dag}(\br,t) \hat{L}^{-1} \mathbf{\Delta}_{\cK}(\br,t) \right],
\end{align}
where $\mathbf{\Delta}=(\Delta^{\mathrm{cl}},\Delta^{q})^T$. The structure of the fluctuation propagator $\hat{L}$ is characteristic for bosons:
\begin{align}
\hat{L}^{-1}&=\left(\begin{array}{cc}
0 & L^{-1}_A \\ L^{-1}_R & L^{-1}_K
\end{array}\right).
\end{align}
The indexes $R$, $A$ and $K$ denote the retarded, advanced and Keldysh propagators, respectively. The cl-cl component of $\hat{L}$ is zero.
This is expected, since for $\Delta^q=0$ the field on the upper and lower branches of the Keldysh contour is the same, and therefore the
corresponding actions cancel each other, resulting in $S_{\mathrm{\GL}}(\Delta^{cl},0)=0$.
The Keldysh propagator is responsible for the Gaussian noise term in the time-dependent GL (TDGL) equation,
and also, it carries information about distribution of electrons and holes.
The details of the derivation, as well as general formulas for the fluctuating propagators (that are valid if the system is far from the transition)
are presented in the App.~\ref{app:GL}.
Close to the transition into superconducting state, the propagators become
\begin{align}
L^{-1}_{K}&=i\frac{\pi}{2}\left(1-\tilde{F}^h_0\tilde{F}^e_0\right),\\\label{eq:L_R}
L^{-1}_{R/A}&=\frac{\pi}{8 T_e}\Big\{-(\tau_{\mathrm{\GL}}z_{\cp})^{-1}\notag\\ &\phantom{aw}+
\Big[\mp 4iT_e \tilde{F}^R_{0}+D(\bnabla-2ie\mathbf{A}_{\cK}^{\mathrm{cl}})^2\mp \partial_{t}\mp 2ie\phi_{\cK}^{\mathrm{cl}}
\Big]\notag
\\ &\phantom{aw}\times \Big(1\pm i\frac{T_e}{\Omega}\Big)\Big\}.
\end{align}
Here $\tilde{F}^{e/h}_{\epsilon}=F^{e/h}_{\epsilon\pm e\phi_{\cK}^{\mathrm{cl}}}$ denotes the gauge invariant distribution function,
while other parameters appearing in the retarded and advanced propagators are the functionals of  $F^R_{\epsilon}=(F^h_{\epsilon}-F^e_{-\epsilon})/2$:
\begin{align}\label{eq:Te}
T_e^{-1}&=2\frac{\dif \tilde{F}^R_{\epsilon}}{\dif \epsilon}\Big|_{\epsilon=0},\\
\Omega^{-1}&=\frac{2}{\pi}\dashint \dif \epsilon\frac{\tilde{F}^R_{\epsilon}-\tilde{F}^R_{0}}{\epsilon^2},\\\label{eq:z}
z_{\cp}^{-1}&=1+\left(\frac{T_e}{\Omega}\right)^2.
\end{align}
They are strongly drive dependent, as will be demonstrated in Sec.~\ref{Sec:example}.
The symbol $\dashint$ denotes the principal value of the integral.
Note that the existence of effective temperature $T_e$, does not imply the local equilibrium form of excitation distribution functions.

The nonequilibrium GL relaxation rate is defined as
\begin{align}\label{eq:GLrate}
\tau_{\mathrm{\GL}}^{-1}=&-\frac{4}{\pi}z_{\cp}T_e\int_{-\omega_D}^{+\omega_D} \dif\epsilon \frac{\tilde{F}^R_{\epsilon}- \tanh{\left(\frac{\epsilon}{2T_c}\right)}}{\epsilon}
\notag\\
&+4z_{\cp}\frac{T_e^2}{\Omega}\tilde{F}^R_0\,,
\end{align}
where $\omega_D$ is the Debye energy. The GL relaxation rate represents the inverse lifetime of Cooper pairs, and, therefore, it vanishes
at the transition to the superconducting state.
In an equilibrium, the density matrix is parametrized by the temperature,
and the condition $\tau_{\mathrm{\GL}}^{-1}=0$ tells us that the transition occurs at $T=T_c$.
In a general nonequilibrium case, additional parameters may appear, e.~g.~voltage drop and temperatures of
the thermal baths that are in contact with the system.
Then, the condition $\tau_{\mathrm{\GL}}^{-1}=0$ defines the phase transition surface in the parameter space. We find that any distribution function $F_{\epsilon}$ at the phase transition surface satisfies:
\begin{align}
\int_{-\omega_D}^{+\omega_D}\dif\epsilon \frac{\tilde{F}^R_{\epsilon}- \tanh{\left(\frac{\epsilon}{2T_c}\right)}}{\epsilon}-\pi \frac{T_e}{\Omega}\tilde{F}^R_0=0.
\end{align}
Tuning the external electromagnetic fields and/or temperature of thermal baths, one can control the distance from the phase transition surface. We point out that the theory presented above is valid only for the systems (in the normal state) close to the transition, i.~e.~when $(\tau_{\mathrm{\GL}}T_e)^{-1}\ll 1$. The difference between the Eqs.(\ref{eq:L_R},\ref{eq:GLrate}) and the work in Ref.~\onlinecite{NikolayEPL} is the appearance of $\Omega^{-1}$ and presence of terms $\sim\tilde{F}^R_0$ [results of Ref.\onlinecite{NikolayEPL} are restricted to the nonequilibrium states generated by the voltage between the leads or gates in a certain way]. In general, $\Omega^{-1}$ is nonzero, as discussed in Sec.~\ref{Sec:example}.

In an equilibrium the following relations hold: $F^{e/h}=\tanh{\left(\frac{\epsilon}{2T}\right)}$, $T_e=T$, $\Omega=0$, $z_{\cp}=1$ and we reproduce the standard GL action. It is given by Eq.~(\ref{GLaction}), but with the propagators:
\begin{align}
(L^{-1}_{eq})_{K}&=i\frac{\pi}{2},\\
(L^{-1}_{eq})_{R/A}&=\frac{\pi}{8 T}\left[-(\tau_{\mathrm{\GL}}^{eq})^{-1}+D\bnabla^2\mp \partial_{t}\right],\\\label{eq:tauGL}
(\tau_{\mathrm{\GL}}^{eq})^{-1}&=\frac{8}{\pi}(T-T_c).
\end{align}
In an equilibrium, the Keldysh fluctuation propagator satisfies the relation following from the fluctuation-dissipation theorem
\begin{align}
(L^{-1}_{eq})_{K}=\coth{\left(\frac{\omega}{2T}\right)}\left[ (L^{-1}_{eq})_R(\bq,\omega)-(L^{-1}_{eq})_A(\bq,\omega) \right],
\end{align}
for $\omega\ll T$. We find that in a general case, this relation is violated for a system out of the equilibrium.

Next, we derive the TDGL equation. After the term $\sim|\Delta^q|^2$ in Eq.~(\ref{GLaction}) is decoupled by introduction of the Hubbard-Stratonovich field $\zeta(\br,t)$, one differentiates (\ref{GLaction}) with respect to $\Delta^{q*}$ and obtains:
\begin{align}\label{eq:TDGL}
\frac{8T_e}{\pi}L_R^{-1}\Delta^{cl}_{\cK}+\zeta&=0,
\\ \langle \zeta(\br,t)\zeta^*(\br',t')\rangle&=\frac{16}{\pi\nu} T_e^2\left(1-\tilde{F}^h_0\tilde{F}^e_0\right)\delta(\br-\br')\delta(t-t').
\end{align}
The field $\zeta$ has a meaning of the Gaussian noise. The superconducting order parameter changes under a gauge transformation as $\Delta^{cl}_{\cK}=\Delta^{cl}e^{2ie k^{cl}}$. However, the TDGL equation  for the order parameter is gauge invariant, since the presence of the gauge field $k^{cl}$ in $L^{-1}_R$ (through $\mathbf{A}_{\cK}^{\mathrm{cl}}$ and $\phi_{\cK}^{\mathrm{cl}}$) compensate the change of the order parameter phase.

Using the Keldysh formalism, derivation of the GL close-to-equilibrium action was performed in Ref.\onlinecite{L_Kamenev}. There was utilized a special gauge where $D \bnabla \mathbf{A}_{\cal{K}}+\phi_{\cal{K}}=0$ and found that the scalar potential does not appear in TDGL equation for the order parameter. In Ref.\onlinecite{L_Kamenev} it was used $F_{\epsilon}=\tanh({\epsilon}/{(2T)})$. By looking at kinetic equation (\ref{eq:kinetic}) it becomes obvious that mentioned gauge may indeed simplify the calculation. However, in the presence of external potentials $\tanh({\epsilon}/{(2T)})$ is not the solution of the kinetic equation even in this gauge, due to boundary conditions. As a result, the scalar electromagnetic  potential does appear in the retarded propagator, (\ref{eq:L_R}), and therefore also appears in the TDGL equation for the order parameter, \eqref{eq:TDGL}, contrary to the statements in Ref.\onlinecite{L_Kamenev}.

Although the higher order terms in the superconducting order parameter in the GL action are unimportant for
our analysis of superconducting fluctuations, we will state them for completeness.
We focus only on local fourth order terms of the form $\Delta^{q*}\Delta^{cl}|\Delta^{cl}|^2$ and find:
\begin{align}\label{eq:S4}
S_{\Delta^4}=&-\pi \nu \int_{0}^{\infty}\dif\tau \tau \tilde{Y}(\tau) \notag\\ &\times\Tr\Big[\Delta^{q*}_{\cK}(\br,t)\Delta^{cl}_{\cK}(\br,t)\big|\Delta^{cl}_{\cK}(\br,t) \big|^2\Big] +c.c.,\\
\tilde{Y}(\tau)=&\int_{-\infty}^{\infty}\frac{\dif\epsilon}{2\pi}\frac{\tilde{F}^R_{\epsilon}} {\epsilon+i 0}e^{i\epsilon\tau}.
\end{align}
In the equilibrium the action reduces to  \cite{L_Kamenev} $S_{\Delta^4}=-7\nu\zeta(3)\Tr\Big[\Delta^{q*}_{\cK}(\br,t)\Delta^{cl}_{\cK}(\br,t)\big|\Delta^{cl}_{\cK}(\br,t) \big|^2+c.c.\Big]/(4\pi^2T^2)$.

\section{Corrections to electrical conductivity \label{sec4}}

In this section, we derive corrections to the Drude conductivity arising in thin films due to superconducting fluctuations. The system is close to the transition, but under FFE conditions.

First, we briefly explain the origin of different fluctuation induced contributions.
Fluctuation Cooper pairs carry charge and directly contribute to the electrical conductivity, determining Aslamazov-Larkin (AL) contribution\cite{Aslamazov1,Aslamazov2}.
Since quasi-particles are involved in Cooper pairing, effectively the number of carriers participating to the single-electron charge transfer is decreased.
The fluctuation pseudo-gap opens at the Fermi level in the single-particle spectrum, and results in diminishing of the Drude conductivity for
the so-called density of states (DOS) correction\cite{Varlamovbook}.
The third contribution to conductivity, the (anomalous) Maki-Thompson\cite{Maki,Thompson} (MT) correction is of purely quantum origin.
It arises due to coherent scattering of the electrons, forming a fluctuation Coper pair, on the elastic impurities.
As a result, the diffusion coefficient changes and therefore the Drude conductivity is influenced.

We are interested in linear response to the in-plane electric field, while the system is strongly driven out
of equilibrium due to contacts with thermal baths at different temperatures between which it is sandwiched,
or due to an electric field perpendicular to the plane. We focus on the regime where
fluctuations can be treated perturbatively, meaning that below derived corrections are small compared to the
Drude conductivity, $\sigmaD$.
We take into account terms linear in the Ginzburg number $G_i=(\nu D d_f)^{-1}$, i.e.~$\delta\sigma/\sigmaD\sim G_i$, where $d_f$ is film thickness.
The in-plane dc conductivity is given by
\begin{align}\label{dc conductivity}
\sigma_{xx}=-\frac{1}{2}\lim_{\Omega\to
0}\left(\frac{1}{\Omega}\frac{\delta^2 Z}{\delta
\mathbf{A}_{{\cK},{x}}^{cl}(\Omega)\delta
\mathbf{A}_{{\cK},x}^{q}(-\Omega)}\right)\Bigg|_{{\bf{A}}_{\cK}^{q}=0},
\end{align}
where $Z$ is the Keldysh partition function, Eq.~(\ref{eq:Z}).
The terms in the action, that are linear or quadratic in the vector potential, arise  from the first term of Eq.~(\ref{q:S_Q}) and read:
\begin{align}\label{eq:S_A}
iS_A&=-\frac{i\pi \nu e D}{2}\Tr\left\{\left[\check{W},\left(\bnabla\check{W}\right)\right] \Si\Ak\right\},\\ \label{eq:S_A2}
iS_{A^2}&=\frac{\pi\nu D e^2}{4}\Tr\Big\{[\Si\Ak,\check{\Lambda}][\Si\Ak,\check{\Lambda}]+2 \Si \Ak\check{\Lambda}\Si\Ak\check{\Lambda}\check{W}^2\notag\\
&\phantom{=\frac{\pi\nu D e^2}{4}\Tr\Big\{}+2\Si\Ak\check{\Lambda}\check{W}\Si\Ak\check{\Lambda}\check{W}\Big\},
\end{align}
respectively. The first term of Eq.~(\ref{eq:S_A2}) does not contain superconducting fluctuations and therefore after differentiation over classical and quantum components of the vector potential and averaging over $\hat{Q}$ and $\Delta$ fluctuations (see Eq.~(\ref{dc conductivity})), it gives the Drude conductivity. The other two terms in (\ref{eq:S_A2}), produce the MT and DOS correction to the conductivity. The part of the action linear in the vector potential, $S_A$, gives the AL correction. The Drude conductivity is
\begin{align}
\sigmaD&=-\frac{\pi \nu D e^2}{4}\lim_{\Omega\to 0}\left\{\frac{1}{\Omega}\Tr\left(\check{\Lambda}_{\epsilon}\hat{\sigma}_x \check{\Lambda}_{\Omega+\epsilon}+\hat{\sigma}_x\check{\Lambda}_{\epsilon} \check{\Lambda}_{-\Omega+\epsilon} \right)\right\}\notag\\
&=-\frac{\pi \nu D e^2}{2}\lim_{\Omega\to 0}\frac{\Tr\left( F^e_{\epsilon-\Omega}-F^e_{\epsilon+\Omega}+
F^h_{\epsilon-\Omega}-F^h_{\epsilon+\Omega}\right)}{\Omega}\notag \\
&=2\nu D e^2.
\end{align}
In Eq.~(\ref{dc conductivity}), all the quantum components of vector potential are set to zero after differentiation.
Therefore, after differentiation with respect to $\mathbf{A}^{cl}$ and $\mathbf{A}^q$, in calculation of $\sigma$ we can use the action (\ref{eq:S_Q})
derived in the previous section. We assume stationary situation ($F^{e/h}_{t,t'}=F^{e/h}_{t-t'}$).
The in-plane classical components of the vector potential can be set to zero, since we are interested in the liner response to the in-plane electric field.
Then, we find the saddle point configuration of $\mathbf{w}$ for the action (\ref{eq:S_Q}) and using the GL-like-action (\ref{GLaction}),
we obtain:
\begin{widetext}
\begin{align}\label{corr1}
&\langle\langle
w_{1,2}({\bf{q}})w_{3,4}^*(-{\bf{q}})\rangle\rangle_{Q,\Delta}=\frac{2
i}{\nu}\delta_{\epsilon_1-\epsilon_2,\epsilon_4-\epsilon_3}\frac{-L^{-1}_K
L_{A,1-2}L_{R,1-2}+F^h_{3}L_{R,1-2}+F^e_1L_{A,1-2}}{\left\{D
\mathbf{q}^2-i(\epsilon_1+\epsilon_2)\right\}\left\{D \mathbf{q}^2-i (\epsilon_3+\epsilon_4)\right\}},\\ \label{corr2}
&\langle\langle
\bar{w}_{1,2}({\bf{q}})\bar{w}_{3,4}^*(-{\bf{q}})\rangle \rangle_{Q,\Delta}=\frac{2
i}{\nu}\delta_{\epsilon_1-\epsilon_2,\epsilon_4-\epsilon_3}\frac{-L^{-1}_K
L_{A,1-2}L_{R,1-2}-F^h_{2}L_{A,1-2}-F^e_4L_{R,1-2}}{\left\{D
\mathbf{q}^2+i(\epsilon_1+\epsilon_2)\right\}\left\{D \mathbf{q}^2+i (\epsilon_3+\epsilon_4)\right\}},
\\ \label{corr3}
&\langle\langle
\bar{w}_{1,2}({\bf{q}})w_{3,4}^*(-{\bf{q}})\rangle\rangle_{Q,\Delta}= \frac{2
i}{\nu}\delta_{\epsilon_1-\epsilon_2,\epsilon_4-\epsilon_3}\frac{L^{-1}_K
L_{A,1-2}L_{R,1-2}+F^h_{2}L_{A,1-2}-F^h_3L_{R,1-2}}{\left\{D
\mathbf{q}^2+i(\epsilon_1+\epsilon_2)\right\}\left\{D \mathbf{q}^2-i (\epsilon_3+\epsilon_4)\right\}},\\ \label{corr4}
&\langle\langle
w_{1,2}({\bf{q}})\bar{w}_{3,4}^*(-{\bf{q}})\rangle\rangle_{Q,\Delta}= \frac{2
i}{\nu}\delta_{\epsilon_1-\epsilon_2,\epsilon_4-\epsilon_3}\frac{L^{-1}_K
L_{A,1-2}L_{R,1-2}+F^e_{4}L_{R,1-2}-F^e_1L_{A,1-2}}{\left\{D
\mathbf{q}^2-i(\epsilon_1+\epsilon_2)\right\}\left\{D \mathbf{q}^2+i (\epsilon_3+\epsilon_4)\right\}}.
\end{align}
\end{widetext}
Here the angular brackets denote averaging over fluctuations of $\check{Q},\Delta^{cl/q}$. $F^{e/h}_i=F^{e/h}_{\epsilon_i}$ and
$w_{i,j}=w_{\epsilon_i,\epsilon_j}$ where $i,j=1 \ldots 4$. Also, $L_{A/R,i-j}=\left(L_{R/A}^{-1}(\bq,\epsilon_i-\epsilon_j) \right)^{-1}$.

Now, we can proceed to calculation of different corrections to the Drude conductivity.
We start with DOS correction:
\begin{widetext}
\begin{align}\label{DOS}
\delta \sigma^{\DOS}=-\frac{\nu D
e^2}{8\pi}
\int_{\mathbf{q},\epsilon_3,\epsilon_4}\Big(&(\partial_{\epsilon_3}F^e_3)\langle\langle
w_{3,4}({\bf{q}})w_{4,3}^*(-{\bf{q}})+
\bar{w}_{3,4}({\bf{q}})\bar{w}_{4,3}^*(-{\bf{q}})\rangle\rangle_{Q,\Delta}
\notag\\
&+(\partial_{\epsilon_3}F^h_3)\langle\langle
w_{3,4}^*({\bf{q}})w_{4,3}(-{\bf{q}})
+\bar{w}_{3,4}^*({\bf{q}})\bar{w}_{4,3}(-{\bf{q}})\rangle\rangle_{Q,\Delta}
\notag\\ &-F^h_4\partial_{\Omega} \langle\langle
w_{3,4}({\bf{q}})w_{4+\Omega,3+\Omega}^*(-{\bf{q}}) \rangle\rangle\big|_{\Omega=0}-F^e_4\partial_{\Omega} \langle\langle
w_{3,4}^*({\bf{q}})w_{4+\Omega,3+\Omega}(-{\bf{q}}) \rangle\rangle\big|_{\Omega=0}\notag\\
&+
F^e_3\partial_{\Omega} \langle\langle
\bar{w}_{3,4}({\bf{q}})\bar{w}_{4-\Omega,3-\Omega}^*(-{\bf{q}})\rangle\rangle
\big|_{\Omega=0}+ F^h_3\partial_{\Omega} \langle\langle
\bar{w}_{3,4}^*({\bf{q}})\bar{w}_{4-\Omega,3-\Omega}(-{\bf{q}})\rangle\rangle
\big|_{\Omega=0}\Big),
\end{align}
\end{widetext}
where
$\int_{\mathbf{q},\epsilon_3,\epsilon_4}=\int{\dif\mathbf{q}}/{(2\pi)^2}\int_{-\infty}^{\infty}\dif
\epsilon_3\dif \epsilon_4$. Note that terms that in the limit $\Omega\to 0$ behave as $1/\Omega$
will be canceled out with similar terms from the other corrections (MT and AL), and that is why they are omitted here.
The main contribution close to the transition reads
\begin{align}\label{eq:mainDOS}
\delta\sigma^{\DOS}\approx -\frac{De^2}{4\pi}\mathrm{Im}\left[ \int_{\epsilon,\omega,\bq}\frac{L^{-1}_K\left( \partial_{\epsilon}F^e_{\epsilon}|L^A_{+}|^2+\partial_{\epsilon}F^h_{\epsilon} |L^A_{-}|^2\right)}{ \left\{D\mathbf{q}^2-i(2\epsilon-\omega)\right\}^2}\right],
\end{align}
where $L^A_{\pm}=L_A(\bq,\pm\omega)$. We obtain the DOS correction in a thin film
\begin{align}\label{eq:DOS}
\delta\sigma^{\DOS}\approx -\frac{7e^2\zeta(3)}{\pi^4 d_f}\frac{T_e T_{\cp}}{T_{\DOS}^2}\log{\left( \frac{T_e}{\tau_{\mathrm{\GL}}^{-1}}\right)},
\end{align}
up to logarithmic accuracy. Here $d_f$ is the film thickness and new characteristic temperatures are
\begin{align}
T_{\cp}&=T_e z_{\cp}(1-\tilde{F}^h_0\tilde{F}^e_0),\\\label{Tdos}
\frac{1}{T_{\DOS}^2}&=-\frac{\pi^2}{7\zeta(3)}\mathrm{Re}\left[\int\dif\epsilon \frac{\partial_{\epsilon}\tilde{F}^L_{\epsilon}}{(\epsilon+i0)^2}\right],
\end{align}
$\tilde{F}_L=(\tilde{F}^e_{\epsilon}+\tilde{F}^h_{\epsilon})/2$.
The important contribution in Eq.~(\ref{eq:mainDOS}) comes from small momenta,
and therefore we safely cut the momentum integration at the upper limit $Dq^2_{max}\sim T_e$.
The main contribution in the DOS, MT and AL corrections comes from frequencies $\omega-2e\phi+\epsilon_0\lesssim \tau^{-1}_{\mathrm{\GL}}$.
Here and in the following we assume that the system is close to the transition, such that characteristic energy  scales of $\tilde{F}^{e/h}_{\epsilon}$ are much greater than $\tau_{\mathrm{\GL}}^{-1}$ and $|\epsilon_0|/2$; $\epsilon_0=-4T_e\tilde{F}_0^R+\tau_{\mathrm{\GL}}^{-1}T_e/\Omega$.
If the system is far from the transition and these conditions are not satisfied, then one can start calculation from  Eq.~(\ref{DOS}) and use general fluctuation propagators (see App.~\ref{app:GL}) that are not restricted to low frequencies.

Next, we focus on the MT correction to the conductivity. It is given by:
\begin{align}\label{MT}
\delta\sigma^{\MT}=&-\frac{\nu D
e^2}{8\pi}\int_{\mathbf{q},\epsilon_2,\epsilon_3}\Big(\partial_{\epsilon_2}F^e_{2}
\langle\langle \bar{w}_{2,3}({\bf{q}})w_{3,2}^*(-{\bf{q}})
\rangle\rangle\notag\\
&+\partial_{\epsilon_2}F^h_{2} \langle\langle
\bar{w}_{2,3}^*({\bf{q}})w_{3,2}(-{\bf{q}})
\rangle\rangle\Big).
\end{align}
The main contribution close to the transition is:
\begin{align}\label{MTmain}
\delta\sigma^{\MT}&\approx \frac{-iDe^2}{4\pi}\int_{\mathbf{q},\epsilon,\omega}L^{-1}_K
\frac{\partial_{\epsilon}F^e_{\epsilon}|L^A_{-}|^2+
\partial_{\epsilon}F^h_{\epsilon}|L^A_{+}|^2} {D^2\bq^4+(2\epsilon+\omega)^2
}\\
\label{mt main}
&\approx-\frac{e^2}{\pi
d_f}{T_{\cp}}{\tau_{\mathrm{\GL}}}\ln{\left(
\frac{\tau_{\mathrm{\GL}}}{\tau_{\phi}} \right)}+\delta\sigma^{\DOS},
\end{align}
where we cut-off the infrared divergency in the momentum integration by introduction of the finite dephasing time $D\bq_{min}\sim \tau_{\phi}$, $\tau_{\mathrm{\GL}}\ll\tau_{\phi}$.\cite{Varlamovbook} There are many phase-breaking sources, such as the electron scattering on phonons or paramagnetic impurities, or superconducting fluctuations.\cite{Varlamovbook} The nonequilibrium conditions may affect also the equilibrium phase breaking time. However, we leave this problem for future studies.
By treating energy $\epsilon$ as a complex number in (\ref{MTmain}), we obtain the first term in Eq.~(\ref{mt main}) from the poles of the integrand determined by zeros of the denominator. The second term in Eq.~(\ref{mt main}) comes from the poles of distribution functions $\tilde{F}^{e/h}$. Note that the first term is positive, while the second one is negative. Then, the DOS correction is effectively doubled, although the first term in $\delta\sigma^{\MT}$ is the dominant one close to the transition.

Next we calculate the AL correction to the conductivity:
\begin{widetext}
\begin{align}\label{AL}
\delta \sigma^{\AL}&=-\frac{1}{2}\lim_{\Omega\to 0}\left[
\frac{1}{\Omega}\left\langle\left\langle\frac{\delta (iS_A) }{\delta
A_{x}^{cl}(\Omega)} \frac{\delta (iS_A) }{\delta
A_{x}^{q}(-\Omega)} \right\rangle\right\rangle_{Q,\Delta}\right]\\
&=- \frac{(\pi \nu
De)^2}{2(2\pi)^4}\lim_{\Omega\to
0}\Big\{\frac{1}{\Omega}\int_{\br_1,\br_2,\epsilon_1,\epsilon_2,\epsilon_3,\epsilon_4}
\big\langle\big\langle
\left[w_{1,2}(\br_1)\nabla_{x} w_{2,1+\Omega}^*(\br_1)-
w_{1,2}^*(\br_1)\nabla_{x} w_{2,1+\Omega}(\br_1)
+w\to\bar{w}\right]\notag\\
&\phantom{=- \frac{(\pi \nu
De)^2}{2(2\pi)^4}\lim_{\Omega\to
0}\frac{1}{\Omega}\int_{\br_1,\br_2,\epsilon_1,\epsilon_2,\epsilon_3,\epsilon_4}
}\times\Big[-F_{3-\Omega}^ew_{3,4}(\br_2)\nabla_{x}
w_{4,3-\Omega}^*(\br_2)+F_{3-\Omega}^hw_{3,4}^*(\br_2)\nabla_{x}
w_{4,3-\Omega}(\br_2)\notag\\&\phantom{a=- \frac{(\pi \nu
De)^2}{2(2\pi)^4}\lim_{\Omega\to
0}\frac{1}{\Omega}\int_{\br_1,\br_2,\epsilon_1,\epsilon_2,\epsilon_3,\epsilon_4}
}+F_{3}^e\bar{w}_{3,4}(\br_2)\nabla_{x}
\bar{w}_{4,3-\Omega}^*(\br_2)-F_{3}^h\bar{w}_{3,4}^*(\br_2)\nabla_{x}
\bar{w}_{4,3-\Omega}(\br_2)\Big]
\big\rangle\big\rangle_{Q,\Delta}\Big\}
\end{align}
\end{widetext}
We find that close to the transition the main contribution assumes the form
\begin{align}
\delta\sigma^{\AL}\approx &\frac{(eD)^2\pi}{16 d_f}\left( 1+i\frac{T_e}{\Omega}\right) \frac{1}{T_{\AL}^2}\notag \\ &\times\int_{\bq,\omega}q^2 L^{-1}_K |L_A(\bq,\omega-\epsilon_0)|^2\frac{\partial}{\partial_{\omega}} \left[L_{R}(\bq,\omega-\epsilon_0)\right],
\end{align}
where the new characteristic temperature is given by
\begin{align}\label{eq:TAL}
\frac{1}{T_{\AL}^2}&=\frac{4z_{\cp}}{\pi^2}\left( \frac{1}{T_a^2}-\frac{T_e}{\Omega}\frac{1}{T_aT_b} \right),\\\label{Ta}
\frac{1}{T_a}&=\mathrm{Im}\left[\int\dif\epsilon\frac{\tilde{F}^R_{\epsilon}} {(\epsilon-i0)^2}\right],\\\label{Tb}
\frac{1}{T_b}&=\mathrm{Re}\left[\int\dif\epsilon\frac{\tilde{F}^R_{\epsilon}} {(\epsilon-i0)^2}\right].
\end{align}
Performing the remaining integration over $\bq,\omega$, we obtain in the quasi-two-dimensional case:
\begin{align}\label{eq:AL}
\delta\sigma^{\AL}\approx &\frac{e^2}{2\pi d_f}{T_{\cp}}{\tau_{\mathrm{\GL}}}\frac{T_e^2}{T_{\AL}^2}.
\end{align}
We conclude that all the fluctuation-induced corrections to the conductivity behave differently as a function of $\tau_{\mathrm{\GL}}$. For a thin film close to the superconducting transition, the MT is the most important one. Moreover, each correction is parameterized by different combination of the effective temperatures: $T_e$, $T_{\cp}$, $T_{\DOS}$ and $T_{\AL}$. These temperatures are strongly drive-dependent, as will be shown in the next section when considering a concrete example.
In the equilibrium $F^{e/h}_{\epsilon}=\tanh{\epsilon/2T}$, and $T_e=T_{\cp}=T_{\DOS}=T_{\AL}=T$. Then, we reproduce the well-known results for the DOS, MT and AL corrections to the conductivity.

Note that calculation of the DOS, MT and AL corrections in the equilibrium within the Keldysh formalism was done in Ref.~\onlinecite{L_Kamenev}. In their derivation of DOS correction they have missed the last four terms from Eq.~(\ref{DOS}), that give important contribution in the final result. However, after some canceling mistakes they surprisingly obtained the correct final result.

In this section we have focused on the derivation of the fluctuation conductivity corrections close to the transition. They are the most pronounced in that region, but nevertheless they can be still significant also far from the transition.\cite{Varlamovbook} Then, the derived GL-like theory is not applicable. One has to take into account high-frequencies and short-wave contributions in fluctuating propagators. However, this can be done within above developed approach. Namely, the expressions for the corrections given by Eqs.~(\ref{DOS},\ref{MT},\ref{AL}) are valid also far from the transition. Then, in Eqs.~(\ref{corr1}-\ref{corr4}) one has to use the general expressions for the fluctuating propagators, that are given in the App.~\ref{app:GL}.

\section{Experimental realization\label{Sec:example}}

In this section we propose an experimental setup where our predictions could be tested. The setup is shown in Fig.~\ref{device}, where the superconducting film
is sandwiched between the substrate and the gate and is separated from them by the interface barriers with the resistances $R_1$ and $R_2$, respectively.
We consider the stationary situation and assume that the Thouless energy corresponding to diffusion across the film  $E_T^\perp=D/d_f^2$, well exceeds all the effective temperatures. Then current across the interface separating the substrate and film is
\begin{align}
I=\frac{1}{4 e R_1}\int\dif\epsilon\left\{F^e(\epsilon)-F^e_S(\epsilon)-F^h(\epsilon)+F^h_S(\epsilon) \right\},
\end{align}
and a similar equation holds for the interface between the film and the gate [here the subscript $S$ denotes the substrate]. From the continuity equation for the current follows $F^{e/h}(\epsilon)=x F_S^{e/h}(\epsilon)+(1-x)F^{e/h}_{\Gs}(\epsilon)$, where $x=R_2/(R_1+R_2)$. Here $F_S^{e/h}(\epsilon)=\tanh\left({\epsilon}/(2T_1)\right)$ and $F_{\Gs}^{e/h}=\tanh\left({(\epsilon\mp eV_{\Gs})}/(2T_2)\right)$ denote distributions it the substrate and it the gate. Then the gauge invariant distribution in the film assumes the form
\begin{align}
\tilde{F}^{e/h}(\epsilon)=x \tanh{\left[\frac{\epsilon\pm (1-x)eV_{\Gs}}{2T_1}\right]}\noindent\\ +(1-x) \tanh{\left[\frac{\epsilon\mp x eV_G}{2T_2}\right]},
\end{align}
in the case of very resistive interfaces, i.e.~when the resistance of the film can be neglected: $R_{tot}\approx R_1+R_2$.
Next, we calculate parameters appearing in the GL-like action, Eq.~(\ref{GLaction}), and demonstrate that they are strongly drive dependent.
\subsection{Ginzburg-Landau relaxation time}

\begin{figure}
\includegraphics[width=0.95\columnwidth]{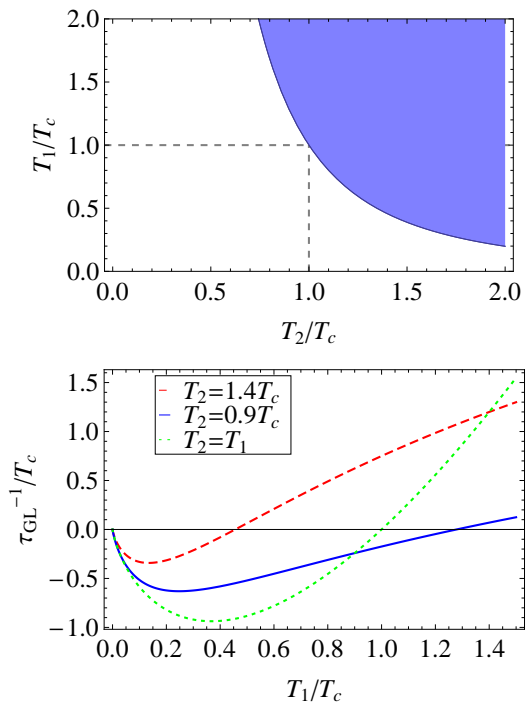}\\
\caption{(Color online) The upper figure shows two regions in which $\tau_{\GL}$ has different signs, for fixed $x=0.3$. In the blue region it is greater than zero, and in the white region it is smaller than zero. The lower figure shows $\tau_{\GL}^{-1}$ dependence on $T_1$, for fixed $T_2$ and $x=0.3$.}\label{fig:zeroV}
\end{figure}
In this subsection we analyze GL relaxation rate under FFE conditions (given by Eq.~(\ref{eq:GLrate})) for the set-up shown in Fig.~\ref{device}. The GL relaxation rate is very important parameter, since many quantities influenced by superconducting fluctuations are singular functions of it. One example are fluctuation induced corrections to the conductivity, that are analyzed in the previous section.
While in the equilibrium ($T_1=T_2$ and $V_{\rm\scriptscriptstyle G}=0$) and close to the transition, it behaves as (\ref{eq:tauGL}), far from equilibrium we find
\begin{align}
\label{eq:Te}
T_\mathrm{e}&=\left[\frac {x}{T_1\ch^2\frac {(1-x)e V_{\rm\scriptscriptstyle G}}{2T_1}}
+\frac {(1-x)}{T_2\ch^2\frac {x e V_{\rm\scriptscriptstyle G}}{2T_2}}\right]^{-1},
\\
\label{eq:invOmega}
\Omega^{-1}&=\frac{2x}{T_1 \pi^2}\mathrm{Im}\left[ \Psi '\left(\frac{1}{2}-i\frac{eV_{\Gs} (1-x)}{2\pi T_1}\right)\right]\notag\\
&+\frac{2(1-x)}{T_2 \pi^2}  \mathrm{Im}\left[\Psi'\left(\frac{1}{2}+i\frac{eV_{\Gs} x}{2\pi T_2}\right)\right],
\\ \label{eq:tGL}\tau_{\mathrm{\GL}}^{-1}&=\frac{8}{\pi}z_{\cp}T_e\Big\{ x\mathrm{Re}\left[\Psi\left(\frac{1}{2}+i\frac{(1-x)eV_{\Gs}}{2\pi T_1}\right)\right]\notag\\& +(1-x)\mathrm{Re}\left[\Psi\left(\frac{1}{2}+i\frac{xeV_{\Gs}}{2\pi T_2}\right)\right]+2\log2\notag\\&+x \log\frac{T_1}{T_2}-\log\frac{T_c}{T_2}+\gamma \Big\}+4z_{\cp}\frac{T_e^2}{\Omega}\tilde{F}^R_0,
\end{align}
where $\Psi(z)$ is the digamma function defined as $\Psi(z)=\Gamma'(z)/\Gamma(z)$, where $\Gamma(z)$ is the gamma function. $\gamma$ is the Euler constant and definitions for $z_{\cp}$ and $\tilde{F}^R$ are given in Sec.\ref{sec:third}. Note that the theory presented in the previous chapters is valid only above ($\tau_{\GL}>0$), and very close to the transition. However, the obtained expression for GL relaxation rate might be valid also below the transition. Also, all the expressions are valid for sufficiently small voltage drop.
\begin{figure}
\includegraphics[width=0.9\columnwidth]{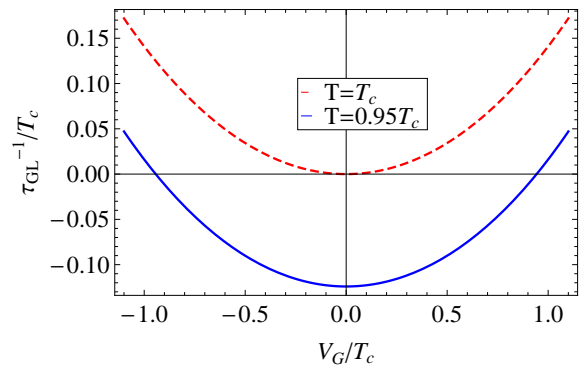}\\
\caption{(Color online) GL relaxation rate dependence on the gate voltage for $T_1=T_2=T$ and $x=0.5$. One distinguishes  quadratic and linear dependence on the gate voltage around the transition for $T=T_c$ and $T=0.95 T_c$, respectively. }\label{fig:nonzeroV}
\end{figure}
The system can be driven from equilibrium due to finite difference of the gate and the substrate temperatures and/or due to gate voltage. First, we start with the zero voltage case. Then the general expressions (\ref{eq:Te}-\ref{eq:tGL}) simplify significantly, and we obtain $\Omega^{-1}=0$, $z=1$ and
\begin{align}
\tau_{\mathrm{\GL}}^{-1}=\frac{8}{\pi}T_e\left(x \log\frac{T_1}{T_2}-\log\frac{T_c}{T_2} \right).
\end{align}
Simple analysis shows that $\tau_{\mathrm{\GL}}^{-1}$ is negative (positive) when both temperatures $T_1$
and $T_2$ are smaller (greater) than the critical temperature $T_c$ and can be either positive or negative when one of the temperatures is greater and another is smaller than $T_c$, see Fig.~\ref{fig:zeroV}. Looking at lower part of Fig.~\ref{fig:zeroV}, one notices that GL relaxation rate can take rather different values than in the equilibrium. The dotted line denotes the equilibrium situation. In the equilibrium, we reproduce the (\ref{eq:tauGL}) when the system is close to the transition ($T-T_c\ll T_c$).
\begin{figure}
\includegraphics[width=0.9\columnwidth]{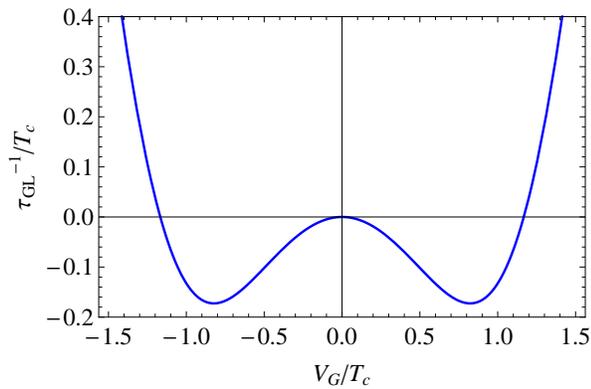}\\
\caption{(Color online) GL relaxation rate dependence on the gate voltage for $x=0.3$, $T_2=2 T_c$ and $T_1=T_2^{((x - 1)/x)} T_c^{(1/x)}\approx 0.2T_c$. \label{fig:question}}
\end{figure}
Next, we switch on the gate voltage. Since the expression for GL relaxation rate is rather complicated, we first analyze the case $T_1=T_2=T$. Then, for $V_{\Gs} \le T$ we obtain
\begin{align}\label{eq:approx}
\tau_{\GL}^{-1}\approx &-\frac{8T}{\pi}\log\frac{T_c}{T}\notag\\&-\frac{2V_{\Gs}^2}{T\pi} x (1-x)\left[ \log\frac{T_c}{T}-\frac{7}{\pi^2}\zeta(3)\right],
\end{align}
where $\zeta(x)$ is Riemann zeta function. Then, one sees that for $T=T_c$ and $V_{\Gs}=0$ the system is at the transition. It moves away from the transition by increasing the gate voltage, as it is shown in Fig.\ref{fig:nonzeroV}. The GL relaxation rate increases quadratically with $V_{\Gs}$ and assumes the form $\tau_{\GL}^{-1}={14V_{\Gs}^2} x (1-x)\zeta(3)/({T_c\pi^3})$. On the other hand, if for zero gate voltage the system is below, but close to the transition ($|\tau_{\GL}^{-1}(V_{\Gs}=0,T)|\ll T<T_c$), then at some critical finite voltage $V_c$ it will be at the transition:
\begin{align}
V_c=T\sqrt{ \frac{4}{x(1-x)}\frac{\log\frac{T_c}{T}} {\frac{7}{\pi^2}\zeta(3)-\log\frac{T_c}{T}}}.
\end{align}
In this case, one obtains linear behavior in $V-V_c$ around the transition, for the fixed temperature: $\tau_{\GL}^{-1}\approx \frac{8}{\pi}\sqrt{\log\left({\frac{T_c}{T}}\right) x (1-x)\left[\frac{7}{\pi^2}\zeta(3)-\log\frac{T_c}{T}\right]} (V-V_c)$. This situation is illustrated in Fig.~\ref{fig:nonzeroV}.
However, for some choice of parameters, the situation can be more complicated, as it is shown in Fig.~{\ref{fig:question}}.
\subsection{Corrections to the electrical conductivity}

\begin{widetext}

\begin{figure}
\includegraphics[width=\columnwidth]{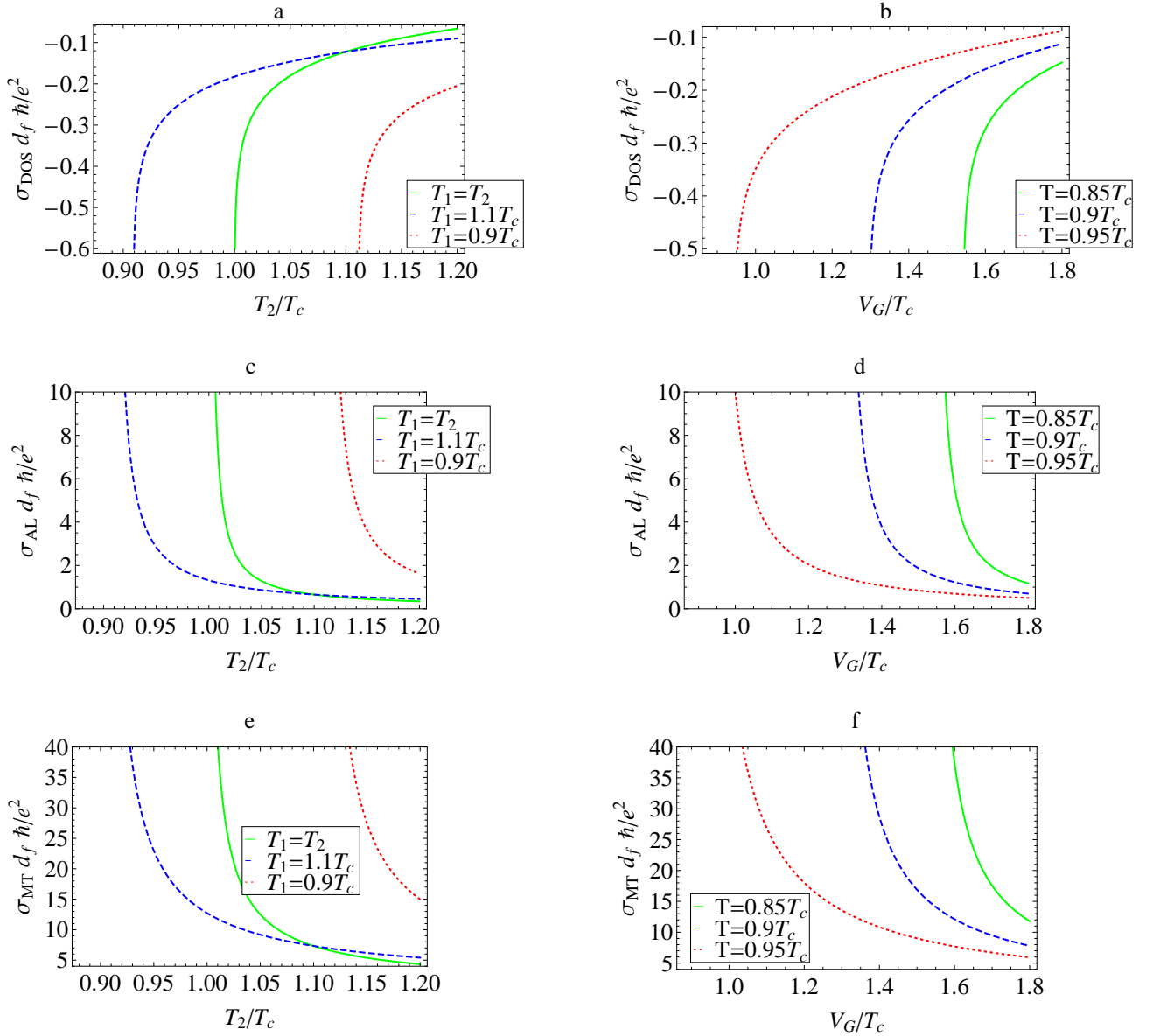}\\
\caption{(Color online) DOS, AL and MT corrections to the in-plane electric conductivity as a function of the gate and the substrate temperatures for zero gate voltage, $x=0.5$ and $\tau_{\phi}^{-1}=10^{-3}T_c$ are shown in Figs.~a, c, e, respectively. DOS, AL and MT corrections to electric conductivity as a function of gate voltage for $T_1=T_2=T$, $x=0.5$ and $\tau_{\phi}^{-1}=10^{-3}T_c$ are shown in Figs.~b, d, f, respectively.}\label{fig:all}
\end{figure}
\end{widetext}

In this subsection we examine fluctuation induced corrections to the in-plane conductivity for the setup in Fig.~\ref{device}. We start with DOS correction, Eq.~(\ref{eq:DOS}). We find that characteristic energy scale $T_{\DOS}$, Eq.~(\ref{Tdos}),  parameterizing DOS correction reads:
\begin{align}
\frac{1}{T_{\DOS}^2}=&-\frac{1}{14\zeta(3)}\frac{x}{T_1^2} \mathrm{Re}\left[\Psi''\left(\frac{1}{2}-\frac{i e V_{\Gs} (1-x)}{2 T_1 \pi}\right)\right]\notag\\&-\frac{1}{14\zeta(3)}\frac{1-x}{T_2^2}\mathrm{Re}\left[ \Psi''\left(\frac{1}{2}-\frac{i e V_{\Gs} x}{2 T_2 \pi}\right)\right].
\end{align}
Now, we have the analytic expression for $\delta\sigma_{\DOS}$ as a function of $T_1$, $T_2$, $V_{\Gs}$ and $x$. In Fig.~\ref{fig:all}a we plot dependence on the gate and substrate temperatures for zero voltage and $x=0.5$. In Fig.~\ref{fig:all}b we plot dependence on gate voltage for $T_1=T_2$ and $x=0.5$.

Next, we analyze AL correction Eq.~(\ref{eq:AL}). We find that the characteristic temperatures $T_a$ and $T_b$, Eqs.~(\ref{Ta},\ref{Tb}), for the given setup become:
\begin{align}
\frac{1}{T_a}=&\mathrm{Re}\left[\frac{x}{T_1\pi}\Psi'\left(\frac{1}{2}+\frac{i eV_{\Gs} (1-x)}{2\pi T_1}\right)\right]\notag\\
&+\mathrm{Re}\left[\frac{1-x}{T_2\pi}\Psi'\left(\frac{1}{2}-\frac{i eV_{\Gs} x}{2\pi T_2}\right)\right],\\
\frac{1}{T_b}=&-\mathrm{Im}\left[\frac{x}{T_1\pi}\Psi'\left(\frac{1}{2}+\frac{i eV_{\Gs} (1-x)}{2\pi T_1}\right)\right]\notag\\
&-\mathrm{Im}\left[\frac{1-x}{T_2\pi}\Psi'\left(\frac{1}{2}-\frac{i eV_{\Gs} x}{2\pi T_2}\right)\right].
\end{align}
Then, we have the analytic form of the AL correction. Its dependence on system parameters is illustrated in Figs.~\ref{fig:all}c, \ref{fig:all}d. All the temperatures that appear in $\delta\sigma_{\MT}$ are already calculated. We plot just the first term in Eq.~(\ref{mt main}), the so-called anomalous part of the MT correction, in Fig.~\ref{fig:all}e and Fig.~\ref{fig:all}f, since the second one is equal to $\delta\sigma_{\DOS}$. The dephasing rate is taken to be $\tau_{\phi}^{-1}=10^{-3}T_c$. Comparing all the corrections to the conductivity, one sees that the most important one close to the transition is the MT correction.

\section{Conclusions and discussion\label{sec:conclusions}}

We have derived GL-like theory valid close to the transition into the superconducting state under FFE conditions. We considered stationary situation and electromagnetic fields slowly varying (with respect to the superconducting order parameter) in space. We found that the parameters appearing in the GL type action are functionals of electron and hole nonequilibrium distribution functions.
Close to an equilibrium, we reproduced known results and  showed that the time-dependent GL equation, that is frequently found in literature\cite{Tinkham, Varlamovbook} is correct, contrary to the findings in Ref.~\cite{L_Kamenev}.

Utilizing the theory, we studied the influence of fluctuations on the electrical conductivity, in FFE situation. We demonstrated that different fluctuation induced corrections are controlled by different effective temperatures. These temperatures are drive dependent and carry information about temperatures and electrochemical potentials of the reservoirs that are in contact with the system. We proposed the experimental setup where our predictions could be tested, see Fig.~\ref{device}.

The approach developed in the present paper allows to analytically treat many other important questions, for example, the influence of superconducting fluctuations on the thermal conductivity under FFE conditions. This question, close to an equilibrium, was a controversial and puzzling issue for a long time (see Ref.~\cite{Niven} and references therein). Hopefully, the final solution is that singular contributions of the the DOS and MT correction to the thermal conductivity cancel each other, while the AL contribution is finite. \cite{Niven} Therefore, experimentally detected structure in the thermal conductivity \cite{exp0,exp1,exp2,exp3}, that previously was believed to be explained by these corrections, needs a new explanation. However, since we have found that different corrections to the electrical conductivity are characterized by different effective temperatures under FFE, it is likely that this is the case also with thermal conductivity. Then, the MT and DOS corrections do not cancel each other but instead produce nontrivial result, that might explain the experiments.

\begin{widetext}

\section{Acknowledgments}

This work was supported by the ANR grant 09-BLAN-0097-01/2, the Russian Foundation for Basic Research (Grant No. 10-02-00700), by the President of the Russian Federation (Grant No. MK-7674.2010.2), the Russian Academy of Sciences programs and by the U.S. Department of Energy Office of Science
through the contract DE-AC02-06CH11357.

\appendix
\section{Ginzburg-Landau action\label{app:GL}}

In this appendix we present a detailed derivation of the nonequilibrium GL-like action Eq.~(\ref{GLaction}) valid in the normal state, but very close to the superconducting transition.

We start from Eq.~(\ref{GL}). In the following, we consider stationary distribution functions and stationary electromagnetic fields, slowly varying in space with respect to the fluctuating superconducting order parameter $\Delta^{cl}$. Then,
\begin{align}
\langle w_{\epsilon_1,\epsilon_2}(\mathbf{q_1})w^*_{\epsilon_3,\epsilon_4}(\mathbf{q_2}) \rangle_{iS_{w^2}}=\frac{2 (2\pi)^{d+2}}{\pi \nu}\frac{\delta(\epsilon_1-\epsilon_4) \delta(\epsilon_3-\epsilon_2)\delta(\mathbf{q}_1+\mathbf{q}_2)}{D(\mathbf{q_1}-2 e \mathbf{A}^{cl}_{\cK})^2-i(\epsilon_2+\epsilon_4)},\\
\langle\bar{w}_{\epsilon_1,\epsilon_2}(\mathbf{q_1}) \bar{w}^*_{\epsilon_3,\epsilon_4}(\mathbf{q_2}) \rangle_{iS_{w^2}}=\frac{2 (2\pi)^{d+2}}{\pi \nu}\frac{\delta(\epsilon_1-\epsilon_4) \delta(\epsilon_3-\epsilon_2)\delta(\mathbf{q}_1+\mathbf{q}_2)}{D(\mathbf{q_1}-2 e \mathbf{A}^{cl}_{\cK})^2+i(\epsilon_2+\epsilon_4)},
\end{align}
where $d$ denotes the dimension. The average values of all other two field combinations give zero contribution. Then, after some algebra we obtain
\begin{align}\label{eq:GLappendix}
S_{\GL}&=2\nu \int \frac{\dif\omega}{2\pi} \frac{\dif\mathbf{q}}{(2\pi)^d}\left[{\Delta}_{\cK,-}^{q*}L_R^{-1} {\Delta}^{cl}_{\cK,+}+{\Delta}_{\cK,-}^{cl*}L_A^{-1} {\Delta}^{q}_{\cK,+}+{\Delta}_{\cK,-}^{q*}L_K^{-1} {\Delta}^{q}_{\cK,+} \right],
\end{align}
where $\Delta_{\cK,\mp}=\Delta_{\cK}(\mp\mathbf{q},\mp\omega)$. The general formula for the retarded part of superconductive fluctuation propagator is
\begin{align}\label{eq:A4}
L^{-1}_R(\bq,\omega)=-\frac{1}{\lambda}-i\int_{-\omega_D}^{\omega_D}\dif\epsilon \frac{F^R_{\epsilon_-}}{D(\mathbf{q}-2 e \mathbf{A}^{cl}_{\cK})^2-2i\epsilon},
\end{align}
where $\epsilon_-=\epsilon-\omega/2$, $F^R_{\epsilon}=(F^h_{\epsilon}-F^e_{-\epsilon})/2$, and $(L^{-1}_A(\bq,\omega))^*=L^{-1}_R(\bq,\omega)$. In the following we show that $L^{-1}_R$ can be written in the form of Eq.~(\ref{eq:L_R}). Taking into account that gauge invariant distribution function is $\tilde{F}^{e/h}_{\epsilon}=F^{e/h}_{\epsilon\pm e\phi_{\cK}^{\mathrm{cl}}}$ and by replacing $\varepsilon=\epsilon_-+e\phi_{\cK}$, one obtains
\begin{align}\label{eq: LR-derivation}
L^{-1}_R=-\frac{1}{\lambda}+ \int_{-\omega_D}^{\omega_D} \dif \varepsilon \frac{\tilde{F}^R_{\varepsilon}-\tilde{F}^R_{0}}{2\varepsilon}
-i\int_{-\omega_D}^{\omega_D}\dif\varepsilon \Big[ \frac{\tilde{F}^R_{\varepsilon}}{D(\mathbf{q}-2 e \mathbf{A}^{cl}_{\cK})^2-2i\varepsilon-i\omega+2 i e \phi_{\cK}^{\mathrm{cl}}} +\frac{\tilde{F}^R_{\varepsilon}-\tilde{F}^R_{0}}{2i\varepsilon}\Big].
\end{align}
Here we have taken into account that Debye frequency $\omega_D\gg \omega/2-e\phi_{\cK}^{\mathrm{cl}}$. Also we have added and subtracted the term $\int_{-\omega_D}^{\omega_D} \dif \varepsilon \left({\tilde{F}^R_{\varepsilon}-\tilde{F}^R_{0}}\right)/{(2\varepsilon)}$ from Eq.~(\ref{eq:A4}). Let us concentrate on last two terms in Eq.~(\ref{eq: LR-derivation}) and denote their sum as $l_R^{-1}$:
\begin{align}\label{eq:lR}
l_R^{-1}=\frac{-i}{4}\int_{-\infty}^{\infty}\dif\varepsilon\left\{ \frac{(\tilde{F}^R_{\varepsilon}-\tilde{F}^R_0)[D(\mathbf{q}-2 e \mathbf{A}^{cl}_{\cK})^2-i\omega-2 i e \phi_{\cK}^{\mathrm{cl}}]}{\varepsilon [\varepsilon + i D(\mathbf{q}-2 e \mathbf{A}^{cl}_{\cK})^2/2+\omega/2- e \phi_{\cK}^{\mathrm{cl}}]}+\frac{2i \tilde{F}^R_0}{\varepsilon + i D(\mathbf{q}-2 e \mathbf{A}^{cl}_{\cK})^2/2+\omega/2- e \phi_{\cK}^{\mathrm{cl}}}\right\}.
\end{align}
The value of the first term in Eq.~(\ref{eq:lR}) is determined by the poles of the function $\tilde{F}^R_{\varepsilon}$ with positive imaginary part. Then, very close to the superconductor-metal transition it becomes
\begin{align}
l_R^{-1}=\frac{-i}{4}[D(\mathbf{q}-2 e \mathbf{A}^{cl}_{\cK})^2-i\omega+2 i e \phi_{\cK}^{\mathrm{cl}}]\int_{-\infty}^{\infty}\dif\varepsilon \frac{(\tilde{F}^R_{\varepsilon}-\tilde{F}^R_0)}{\varepsilon (\varepsilon + i 0)}-\frac{i\pi\tilde{F}^R_0}{2},
\end{align}
since $\omega/2+e\phi_{\cK}^{\mathrm{cl}}$, as well as $D(\mathbf{q}-2 e \mathbf{A}^{cl}_{\cK})^2$, is much smaller than any relevant scale of the distribution functions.
Taking into account that $\int_{-\omega_D}^{\omega_D}\dif\varepsilon \tanh{(\varepsilon/2T)}/\varepsilon\approx 2\log(4\omega_D e^{\gamma}/2\pi T)$ for $\omega_D\gg T$ where $\gamma$ is Euler constant, $T_c=2\omega_De^{\gamma-1/\lambda}/\pi$ and Sokhotsky's formula $(\varepsilon +i0)^{-1}=-i\pi\delta(\varepsilon) + \mathcal{P} (\varepsilon^{-1})$ one obtains:
\begin{align}
L^{-1}_R=\int_{-\omega_D}^{\omega_D}\dif\varepsilon \frac{\tilde{F}^R_{\varepsilon}-\tanh{\left( \frac{\varepsilon}{2T_c}\right)}}{2\varepsilon}+[-D(\mathbf{q}-2 e \mathbf{A}^{cl}_{\cK})^2+i\omega-2 i e \phi_{\cK}^{\mathrm{cl}}]\left(+\frac{\pi}{4}\frac{\dif \tilde{F}^R}{\dif \varepsilon}\Big|_{0}+i\dashint \dif\varepsilon \frac{\tilde{F}^R_{\varepsilon}-\tilde{F}^R_0}{4\varepsilon^2} \right)-\frac{i\pi \tilde{F}^R_0}{2}.
\end{align}
After introducing $T_e$, $\Omega$ and $\tau_{\mathrm{\GL}}$ as given by Eqs.(\ref{eq:Te}-\ref{eq:z}), one arrives at Eq.~(\ref{eq:L_R}).

\end{widetext}
Note that the term $\sim \Delta^{cl*}\Delta^{cl}$ vanishes. This is expected property of the action (see the explanation in the main text). Taking into account that $\partial_t$ in Eq.\ref{eq:S} is just a symbol standing instead of a matrix in the discrete time space, and that sum of the retarded and the advanced Green's functions taken at the same time vanishes, we find that term $\sim\Delta^{cl*}\Delta^{cl}$
is zero, while
\begin{align}
L^{-1}_K=i\int \dif\epsilon \frac{1-\frac{1}{2}\left(F^h_{\epsilon_-}F^e_{\epsilon_+} +F^h_{-\epsilon_+}F^e_{-\epsilon_-}\right)}{D(\mathbf{q}-2 e \mathbf{A}^{cl}_{\cK})^2-2i\epsilon}.
\end{align}
Very close to the transition, it reduces to
\begin{align}
L^{-1}_K&= -\frac{1}{2}\int \dif\epsilon \frac{1-\frac{1}{2}\left(F^h_{\epsilon_-}F^e_{\epsilon_+} +F^h_{-\epsilon_+}F^e_{-\epsilon_-}\right)}{\epsilon+i0}\\&= \frac{i\pi}{2} \left(1-\tilde{F}^h_{-\omega/2+e\phi_{\cK}^{\mathrm{cl}}} \tilde{F}^e_{\omega/2-e\phi_{\cK}^{\mathrm{cl}}}\right)\\ &\approx \frac{i\pi}{2} \left(1-\tilde{F}^h_0\tilde{F}^e_0\right).
\end{align}
Note that after applying the Sokhotsky's formula, the term $ \dashint\dif \epsilon (F^h_{\epsilon_-}F^e_{\epsilon_+}+F^h_{-\epsilon_+}F^e_{-\epsilon_-})/\epsilon$ gives zero contribution, since the function under the integral is the odd function of $\epsilon$.

In the above calculation we used that close to the transition $\omega/2-e\phi_{\cK}^{\mathrm{cl}}$, as well as $D(\mathbf{q}-2 e \mathbf{A}^{cl}_{\cK})^2$, is much smaller than any relevant scale of the distribution functions. Let us demonstrate the importance of the mentioned combination of the momenta/frequancy and gauge fields by considering simple example. Let us consider an equilibrium situation and denote the order parameter by $\Delta_0(\omega,\mathbf{q})$ in this case. It is satisfied $D \mathbf{q}^2 \sim \omega \sim (\tau_{\mathrm{\GL}}^{\mathrm{eq}})^{-1}\ll T$. Next we turn on a constant scalar potential $\phi$. Then, the order parameter becomes $\Delta=\Delta_0 \exp{\left(-2ie\phi t \right)}$, i.e.~$\Delta(\omega,\mathbf{q})=\Delta_0(\omega- 2e\phi,\mathbf{q})$. Then, $\omega-2e\phi\sim (\tau_{\mathrm{\GL}}^{\mathrm{eq}})^{-1}\ll T$. Similarly, it can be shown that the combination $D(\mathbf{q}-2 e \mathbf{A}^{cl}_{\cK})^2$ has to be compared with characteristic energies of the distribution function. That is why it is necessary to introduce the gauge invariant distribution functions during the calculation, as we did above.


\end{document}